# ZnCdO:Eu Epitaxially Grown Alloys for Self-Powered Ultrafast Broadband Photodetection

*Igor Perlikowski[1]\*, Eunika Zielony[1], Aleksandra Wierzbicka[2], Anastasiia Lysak[2], Rafał Jakieła[2], Ewa Przeździecka[2]*

[1] Department of Experimental Physics, Wroclaw University of Science and Technology, Wybrzeze Wyspianskiego 27, 50-370 Wroclaw, Poland

[2] Institute of Physics, Polish Academy of Sciences, Al. Lotnikow 32/46, 02-668 Warsaw, Poland

*\* Corresponding author: igor.perlikowski@pwr.edu.pl*

**Abstract**

Photodetectors (PDs) are essential in imaging, communication, and sensing technologies. However, their reliance on external power makes them energy-consuming. This creates a strong need for self-powered PDs as a sustainable alternative. ZnO is a promising semiconductor material due to its pyroelectric properties, stemming from non-centrosymmetric wurtzite crystal structure, enabling the pyro-phototronic effect that enhances response speed. Properties of ZnO can be tailored via alloying and doping. Thus, this work explores thin layers of ZnCdO:Eu random alloys grown by molecular beam epitaxy (MBE) on silicon substrates, with varying Cd content. The study shows that doping with Eu notably affects growth kinetics, promoting strong [0001] orientation preference. Moreover, photoluminescence measurements confirm the successful incorporation of $Eu^{3+}$ ions into the structure. Electrical measurements show that the introduction of Cd eliminates the problem of Schottky barrier formation on the ZnO/Si interface. The n-ZnCdO:Eu/p-Si junctions exhibit rectifying behavior and generate photocurrent across 380–1150 nm wavelength range without external electrical bias. Utilizing the pyro-phototronic effect, these devices achieved ultrafast response times: rise time below 10 µs and decay time below 5 µs for 405 nm and 650 nm illumination – placing them among the fastest self-powered oxide-based detectors that do not rely on additional performance-enhancing layers.

**Introduction**

ZnCdO alloy has been drawing attention of the scientific world for 30 years now.[1] It is a semiconductor material that offers a tunable band gap ranging from 3.3 eV (pure ZnO[2]) to 2.18 eV (pure CdO[3]) at room temperature (*RT*), making it a potential alternative for InGaN in optoelectronic devices.[1] However, as ZnO has a wurtzite structure and CdO crystallizes in a rock salt phase in normal conditions, obtaining good quality structures in case of ZnCdO remains an issue.[1,4,5] Despite that, ZnCdO is being presented as a transparent conductive oxide[1,6] or as a candidate for an active material in devices such as light emitting diodes.[7–9] In the latter application, ZnCdO is often used as a well material in multiple quantum well systems with ZnO serving as the barrier layer.[9,10] Nevertheless, obtaining a sharp interface in these quantum structures is challenging, as Cd demonstrates tendencies to diffuse into adjacent layers.[11]

Recent years have brought data concerning various aspects of ZnCdO. It has been presented as a digital alloy in the form of a short-period {ZnO/CdO} superlattice, offering an alternative to random alloys.[12–15] Additionally, ZnCdO doped with $Co^{2+}$[11] or Ti[16] has been investigated. Regarding doping, pure ZnO doped with various rare-earth elements (REs), such as Eu, La, Tb and Y, has gained popularity.[17,18] In response to this growing interest, our previous research described {ZnCdO/ZnO} superlattices doped with Eu grown on Si.[19]

REs are used in ZnO-based materials to modify optical and electrical properties.[20–22] They can serve as luminescence centers, enabling f-f and f-d internal orbital transitions.[18] Specifically, $Eu^{3+}$ ions exhibit a 612 nm transition, making them suitable for potential red LED applications.[23] A proper concentration of Eu in ZnO has also been reported to induce ferromagnetism at *RT*.[24] Moreover, Eu doping contributes to a slight decrease of the optical band gap,[25] provides charge carriers, reduces resistivity, and enhances the mobility of ZnO films.[26] These last three factors are crucial for solar cells and photodetectors, as they facilitate the escape of photogenerated electrons from the junction into the electrical circuit, increasing photocurrent rather than allowing recombination.

Over the past decade, ZnO has found a specific application in photodetectors (PDs) based on the pyro-phototronic effect.[27–31] This is due to the non-centrosymmetric crystallographic structure of wurtzite ZnO.[32] In such devices, ZnO serves as a source of an additional electric field that is generated when the intensity of incident light changes, causing a temporal variation in temperature. This field can be beneficial in photodetectors as it helps to separate photogenerated carriers, resulting in higher photocurrent[32] and shorter response times.[33] However, most published studies focus on PDs with pure ZnO,[27–31] with only a few

exploring ZnO-based photodetectors doped with elements such as Ga,[34] Fe,[35] and halogens.[36]

In this research, ZnCdO:Eu thin films grown on Si via plasma-assisted molecular beam epitaxy (PA-MBE) are investigated. The introduction of the Eu dopant was performed *in situ* as an alternative to ion implantation,[37–40] a more commonly used method for incorporating REs but one that also caused greater damage to the crystal lattice. The impact of Cd content on lattice strain and vibrational properties is analyzed. We confirm that ZnCdO:Eu/Si system can be utilized as a self-powered broadband photodetector without applying external voltage bias in the range of 380-1150 nm. Under these conditions, photocurrent generation exceeding the level of 400 mA/W for 700 nm is observed, making ZnCdO:Eu thin films highly attractive for energy-saving optoelectronics. Additionally, pyro-phototronic effect was detected, contributing to a fast response of the device. The measured rise times were shorter than 10 μs with fall times below 4 μs, classifying these detectors as ultrafast.[41,42] This research opens a discussion on doped wurtzite-ZnCdO as a potential alternative to pure ZnO in PDs based on pyro-phototronic effect, offering greater flexibility in band gap and strain engineering.

**Experimental Details**

$Zn_{x-1}Cd_xO$ thin films doped *in situ* with Eu were grown on *p*-type Si (001) substrates by plasma-assisted molecular beam epitaxy (Compact 21 Riber). The substrates were degassed in a load chamber at 150°C for 1 hour. Thereafter, the substrate temperature in the growing chamber was raised to 550°C for 10 minutes and then reduced to the growth temperature 380°C. The radio-frequency (RF) cell was used to generate oxygen plasma with a power of 400 W and an oxygen flow of 3 ml/min. The temperature of the Zn and Eu effusion cells was fixed at 573°C (flux: $9.6·10^{-7}$ Torr) and 440°C (flux: $0.4·10^{-9}$ Torr), respectively. Different Cd concentrations in the films were achieved by changing the temperature of cadmium effusion cell (320°C, 330°C and 340°C), and hence the flux ($1.2·10^{-8}$ Torr, $3.0·10^{-8}$ Torr and $3.9·10^{-8}$ Torr, respectively). As a result, four structures were obtained – one ZnO:Eu (EZO) sample and three $Zn_{x-1}Cd_xO$:Eu (CEZO) samples: CEZO320, CEZO330, CEZO340, where the numbers correspond to the temperature of the cadmium effusion cell. Thicknesses of the thin films were ~345 nm, ~335 nm, ~445 nm, ~445 nm for EZO, CEZO320, CEZO330, CEZO340, respectively. Small fragments of each sample underwent rapid thermal processing (RTP) for 5 min at 700°C in oxygen atmosphere. AccuThermo AW610 from Allwin21 Inc. system was used for RTP. To enable electrical measurements, Au contacts were sputtered on the thin films, while Al contacts were deposited on Si substrates.

CAMECA IMS6F system was utilized for secondary ion mass spectroscopy (SIMIS). X-ray diffraction (XRD) measurements were performed using high resolution Panalytical X'Pert Pro MRD diffractometer equipped with $Cu_{K\alpha 1}$ radiation, hybrid 2-bounce Ge (220) monochromator, threefold Ge (220) analyzer in front of proportional detector or Soller slits in front of Pixcel detector. In this work θ/2θ scans were measured in low angle resolution mode (Soller slits and Pixcel detector). Scanning electron microscope Hitachi Su-70 equipped with

Gatan MonoCL3 system and liquid helium cryostat were used to obtain low-temperature (~ 5 K) cathodoluminescence spectra.

To collect Raman spectra, a HORIBA Jobin Yvon T64000 system was used. It was configured for backscattering geometry and operated in a single subtractive mode. The spectrometer was working with 0.1 mm slits, resulting with 0.5 cm$^{-1}$ spectral resolution. The samples were excited by a 532 nm semiconductor laser. A CCD detector was used for scattered light detection. Peak positions of Raman modes were found by fitting the data with Lorentz functions. Raman spectra measurements were repeated 15 times for different spots on each of the samples. Micro-photoluminescence (μPL) was measured using the system described above. However, a 325 nm He-Cd laser was used to excite the structures and slit size was changed to 0.5 mm.

Current-voltage (I-V) characteristics were collected at *RT* using myDLTS software[43] and the following hardware: Keithley 2601A I-V source meter and Zurich Instruments MFIA Impedance Analyzer. During current-time (I-t) characteristics measurements, the samples were illuminated using 405 nm and 650 nm lasers. A Bentham PVE300 Photovoltaic Device Characterization System was used to record responsivity spectra. These results were accompanied by reflectance spectra measured using Jasco V770 UV–visible/NIR spectrophotometer. An I-V curve tracer with a PET Solar Simulator (#SS100AAA) were utilized to collect dark and light I-V data (AM1.5G conditions, 1000W/m$^2$ light intensity, 25°C ambient temperature).

**Results and Discussion**

*Structural properties*

Collected signals for Cd and Eu during SIMS measurements were at the verge of the system detection possibilities. Concentration of Cd reached maximum of $5 \cdot 10^{18}$ cm$^{-3}$ for sample CEZO340 (cf. Figure S1). The results for Eu were not conclusive. However, the presence of Eu$^{3+}$ ions is confirmed later in this section by μPL data.

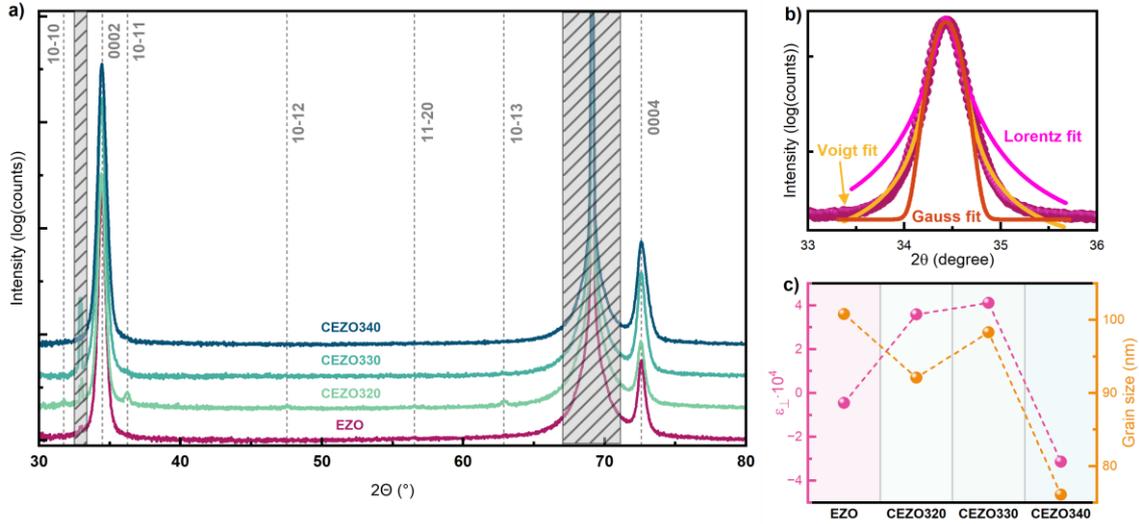

Figure 1. a) X-Ray diffraction patterns for ZnO:Eu and ZnCdO:Eu films in logarithmic scale. Si-related range is hatched. The positions of ZnO-related peaks are marked according to the JCPDC 00–005–0664 card. b) Comparison of fit functions applied to 0002 ZnO peak; c) calculated out-of-plane strain $\varepsilon_\perp$ and grain size based on XRD data – dashed lines added to guide the eye.

X-ray diffraction patterns in a logarithmic scale are shown in Figure 1a. The results confirm growth of wurtzite ZnO. All the spectra feature dominant peaks 0002 ZnO and 0004 ZnO corresponding to the (0001) orientation of the wurtzite ZnO phase. The remaining ZnO-related signals are relatively weak. The XRD scans demonstrate strong preference of ZnO to grow in [0001] direction of wurtzite structure when doped with Eu. Similar observations were made in our previous study regarding Eu-doped {ZnCdO/ZnO} superlattices grown on Si by Lysak et al [44]. The literature describing Eu-doped ZnO films and nanostructures confirms this preference for various substrates.[26,45–49] Additionally, a very sharp peak can be observed at 33° (see Figure 1a). It comes from 200 Si quasi-forbidden reflection - similar results were observed in our XRD measurements of GaN nanowires on Si (100) substrates.[50] The quasi-forbidden reflections are registered because the incident X-ray beam is formed by hybrid 2-bounce monochromator and the beam is divergent (12 arcsec). The existence of quasi-forbidden reflections has been already described by Renninger more than 60 years ago.[51]

To calculate the lattice parameters from XRD scans, the interplanar distance $d_{hkl}$, where $hkl$ denotes the Miller indices, must be determined using Bragg's law[44,52] expressed as:

$$d_{hkl} = \frac{m\lambda}{2\sin\theta_{hkl}}, \qquad (1)$$

where $m$ is the order of reflection ($m$ = 1), $\lambda$ denotes X-ray wavelength ($\lambda$ = 1.54056 Å), and Θ is the Bragg diffraction angle obtained by fitting the peaks with a Voigt function. An example of fitted peak is shown in Figure 1b. As illustrated, neither Gaussian nor Lorentzian functions match the experimental data properly. To better reproduce the shape of the diffraction signal, a Voigt function – being a convolution of both Gaussian and Lorentzian profiles – is employed.

For the wurtzite structure, the relation between $d_{hkl}$ and lattice parameters $a$ and $c$ is determined by the quadratic equation as follows:[44,52]

$$\frac{1}{d_{hkl}^2} = \frac{4}{3} \cdot \frac{h^2+hk+k^2}{a^2} + \frac{l^2}{c^2}. \tag{2}$$

Since the only well-pronounced ZnO-related XRD peaks in all the signals correspond to the [0001] orientation, only the $c$-axis lattice parameter can be found. The peak positions of the ZnO 0002 reflection, with $hkl$ = 002 were used for this analysis. By combining equations (1) and (2) under these conditions, the following relation can be derived:

$$c = \frac{\lambda}{\sin\theta}. \tag{3}$$

Using the obtained $c$ lattice constant, the out-of-plane strain component can be calculated from the equation:[44]

$$\varepsilon_\perp = \frac{c-c_0}{c_0}, \tag{4}$$

where $c_0$ is the reference lattice constant of relaxed ZnO bulk material. The grain size $D$ of ZnCdO:Eu films can be calculated using the Lorentzian component of the diffraction line broadening (FWHM, $\beta_L$ in radians), obtained from fitting the diffraction peaks with Voigt function. The grain size is determined according to the Debye-Scherrer formula:[53,54]

$$D = \frac{k\lambda}{\beta_L \cos\theta}, \tag{5}$$

where $k$ is a constant that equals 0.94 for spherical particles.[54] Parameters calculated based on equations (4) and (5) are depicted in Figure 1c. $Cd^{2+}$ and $Eu^{3+}$ ions have similar ionic radius – 0.097 and 0.095 nm, respectively[55] – both significantly larger than that of $Zn^{2+}$, which totals 0.074 nm.[56] Hence, the modification of strain in the material with doping content is expected. According to our results, sample without Cd displays near-zero strain along the growth direction (out-of-plane). Then, compressive strain increases with Cd content. However, for the sample with the highest Cd concentration, $\varepsilon_\perp$ changes sign and becomes tensile. Calculated grain size is in the range of 75-105 nm.

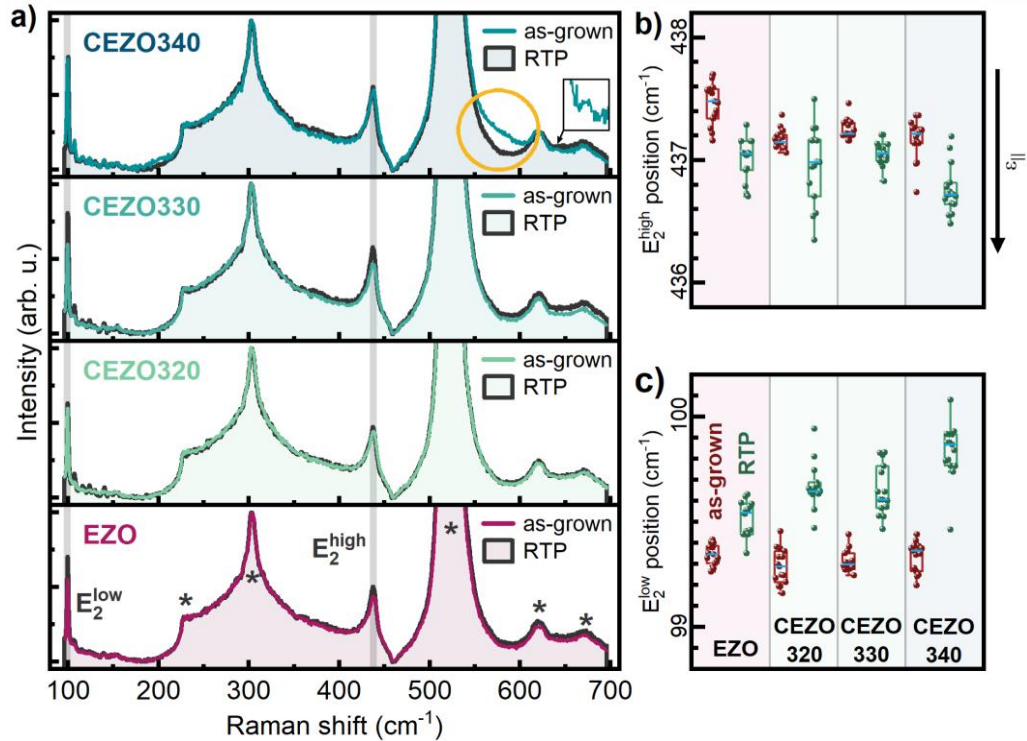

Figure 2. a) Raman spectra collected for as-grown and annealed (RTP) samples. The spectra were normalized to the 303 cm$^{-1}$ Si mode. Comparison of the peak positions of b) $E_2^{high}$ mode with corresponding in-plane strain, $\varepsilon_{||}$, and c) $E_2^{low}$ mode for the studied samples. Red data points correspond to as-grown structures, while green data points represent structures after RTP. Blue lines inside the boxes indicate median values. Each box contains 50% of the data, ranging from the first and third quartile.

Raman spectroscopy was applied to analyze the vibrational properties of ZnCdO:Eu films. In this case, to obtain additional information regarding crystal lattice behavior, small fragments of the samples underwent through rapid thermal processing (RTP). As shown in Figure 2a, the spectra were dominated by Si-related modes, with the 521 cm$^{-1}$ $LTO(\Gamma)$ mode[57] being the most intense. The peaks having origin in the Si substrate are marked with asterisks. A detailed description of the Si bulk Raman spectrum can be found in our previous study.[19] The ZnO $E_2^{low}$ and $E_2^{high}$ modes were found at 99 and 437 cm$^{-1}$, respectively. Contrary to Eu-doped {ZnCdO/ZnO} superlattices,[19] no additional modes were observed except for the CEZO340 as-grown sample. In this case, an asymmetrical broadening of the Si $LTO(\Gamma)$ mode was recorded (Figure 2a, marked with an orange circle). This is probably due to a defect-related band emerging in the ZnO Raman spectra, possibly attributed to the $A_1(LO)$[58] or the mixed $qA(E)_1$ mode.[19,59] The contribution of second-order modes cannot be excluded.[60] The appearance of these Raman modes is a sign of a disordered ZnO lattice, possibly due to zinc interstitials and oxygen vacancies.[58,60,61] Moreover, a weak 640 cm$^{-1}$ peak can be spotted, generally identified as an 'additional mode',[62] or more specifically as the $TA + B_1^{high}$ silent mode.[19,63] This band is commonly observed in ZnO-based structures doped with various elements, such as N, Eu, Ga, and Fe.[19,62,63] Both of these features disappear after annealing, a behavior also observed in Eu-doped {ZnCdO/ZnO} SLs.[19]

An investigation of the shifts of Raman modes can provide insights into the strain present in the structure. The $E_2^{high}$ ZnO mode is known to shift linearly with increasing in-plane strain, $\varepsilon_\parallel$.[19,64] Specifically, a redshift of the $E_2^{high}$ mode from its reference unstrained position indicates tensile strain, whereas a blueshift suggests compressive strain.[19,64] The measured $E_2^{high}$ peak positions are collected in Figure 2b. All the values are below the reference value for unstrained ZnO bulk which is 438.5 cm$^{-1}$ (Raman spectrum not shown here). Hence, tensile in-plane strain is detected in all Zn(Cd)O:Eu films. Moreover, annealing results in a further redshift of the $E_2^{high}$ mode, indicating an increase in strain. This effect was also observed for Eu-doped {ZnCdO/ZnO} SLs.[19] In contrast, $E_2^{low}$ ZnO mode shows the opposite behavior (cf. Figure 2c), where annealing causes a blueshift of this peak.

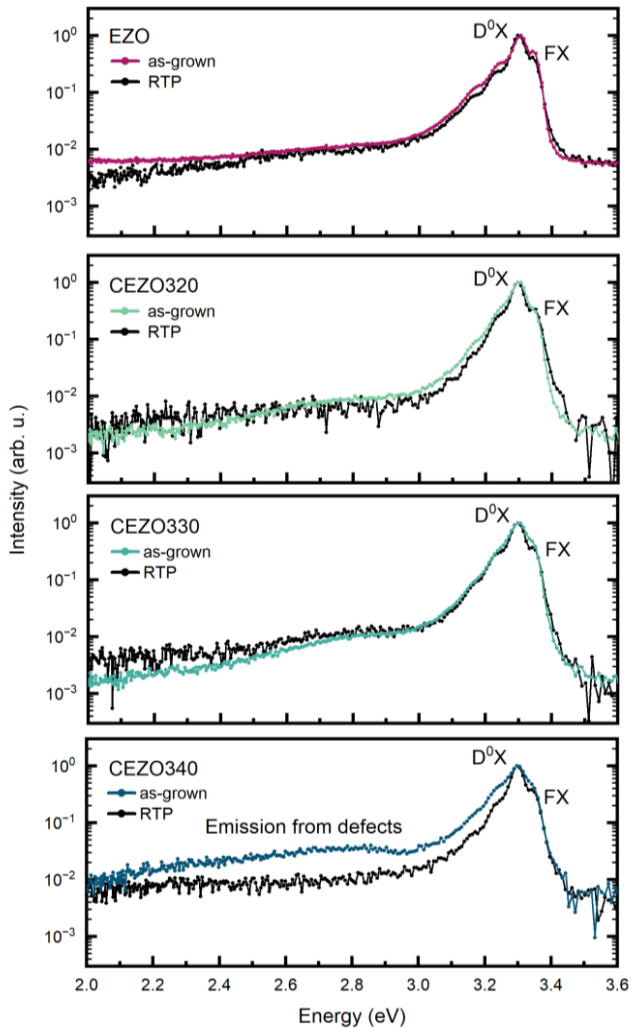

Figure 3. Low-temperature normalized cathodoluminescence spectra of as-grown and annealed samples.

Figure 3 shows cathodoluminescence spectra measured for the as-grown structured and the fragments of the samples annealed in oxygen by RTP method. The most intensive peak detected at about 3.3 eV most probably comes from donor bound exciton states $D^0X$.[65] A less intensive peak related to free exciton (FX) emission is visible at about 3.343 eV as well. Moreover, the defect-related broad band emission can be observed in the 2.0-3.0 eV region (visible spectral range). This band is emerging for both as-grown and annealed samples. In the case of the sample with the highest Cd concentration (CEZO340), a stronger defect-related emission is observed. The origin of this band is most likely a result of native defects in ZnCdO lattice,[66] supporting previous observations from Raman spectroscopy (cf. Figure 2a). Similarly to the results obtained by that method, the CL spectra show that in case of CEZO340 structure RTP reduces the concentration of defects, as evidenced by the disappearance of the broad emission band, as shown in Figure 3.

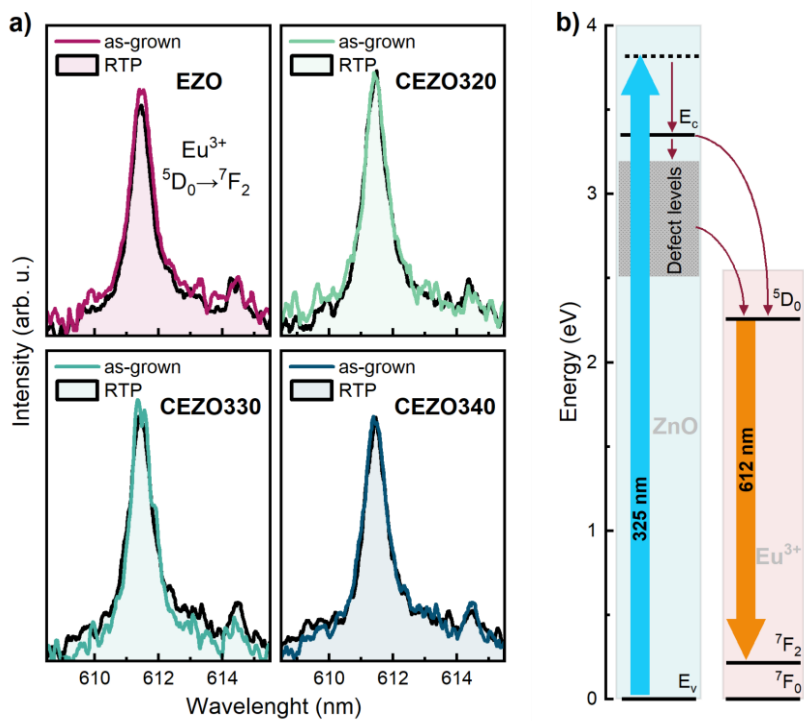

Figure 4. a) Micro-photoluminescence spectra displaying emission from $Eu^{3+}$ ions, and b) schematic representation of the energy transfer process observed during the measurement, based on Zhang *et al*.[67] The samples were excited with a 325 nm laser.

The micro-photoluminescence spectra shown in Figure 4a confirm the successful doping of the material with $Eu^{3+}$ ions, providing additional carriers to the material. The observed emission peak at 612 nm corresponds to the $^5D_0 \rightarrow ^7F_2$ transition of the $Eu^{3+}$ ions. No significant differences were detected in the emission spectrum – neither the Cd content nor the rapid thermal processing influenced the emission from $Eu^{3+}$ ions. Figure 4b represents the possible energy transfer mechanism occurring during the experiment. Upon absorption of 325 nm (3.81 eV) photons, electrons are excited from the valence band ($E_v$) to above the conduction band ($E_c$). Then, they relax to $E_c$. Nonradiative transfer of these electrons from

ZnO to the $^5D_0$ level of $Eu^{3+}$ ions may occur either directly or via defect levels, such as $Eu_{Zn}$ substitutions or zinc interstitials. This is followed by the radiative $^5D_0 \rightarrow ^7F_2$ transition that takes place, which produces the characteristic emission.[67] Besides this peak, a well-pronounced, low-intensity peak at 614.5 nm is observed, likely originating from the $^5D_1 \rightarrow ^7F_4$ transition.[68]

*Electrical properties*

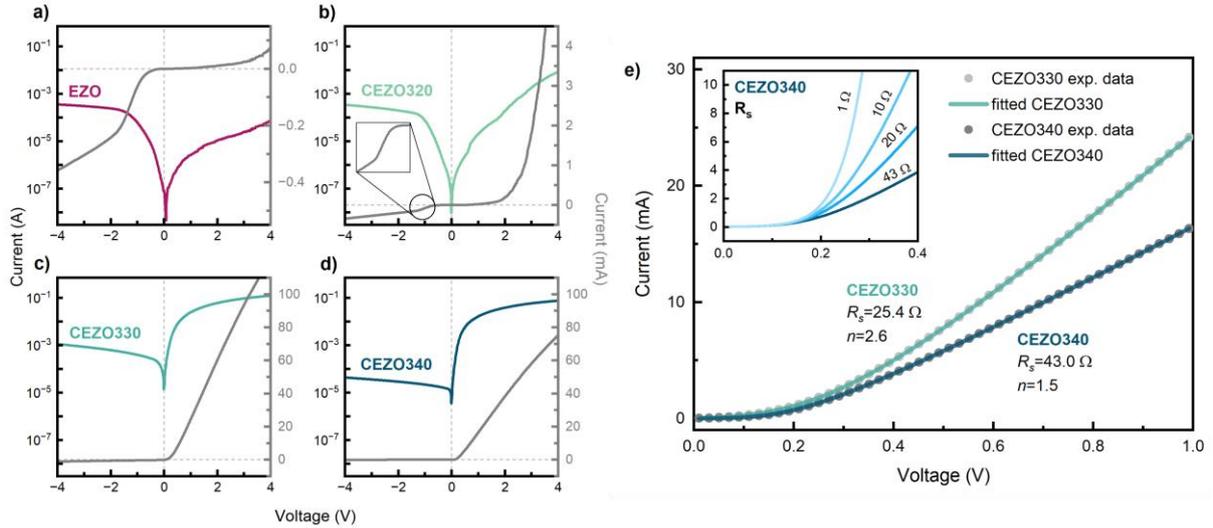

Figure 5. Dark current-voltage curves of samples a) EZO, b) CEZO320, c) CEZO330, and d) CEZO340 measured at room temperature. Colored curves are presented in a semi-logarithmic scale (left axis), whereas the grey curves are on a linear scale (right axis); e) fitted experimental data and the extracted electrical parameters. The inset illustrates the influence of series resistance, $R_s$, on the low-voltage I-V curve of sample CEZO340.

The current-voltage measurements indicate difficulties in obtaining a rectifying junction for structures without Cd and those with the lowest Cd content, namely EZO and CEZO320 (cf. Figure 5a and the inset in Figure 5b, respectively). This is due to the formation of an undesirable potential barrier, evidenced by a sudden increase of current in reverse bias at -1 V. The application of Au contacts directly onto ZnCdO:Eu thin film may contribute to this behavior, as Au can form a Schottky junction with ZnO.[69,70] However, some reports describe devices with Au electrodes applied on ZnO that do not feature this issue.[71,72] The turn-on voltages, $V_{on}$, for these samples were estimated to be at the level of ~ 2.0 V, which aligns with values commonly reported in the literature, typically ranging from 1 to 2.5 V.[73–75] The rectifying factor (RF) at ±4 V totals 0.2 and 25 for EZO and CEZO320, respectively.

Samples 18A and 18B (cf. Figure 5c and d) do not seem to encounter the problem of the additional potential barrier – this is probably due to the changes in, i.a., work function or electron affinity caused by increased Cd concentration.[76] It impacts band alignment leading to ohmic contact formation. RF of CEZO330 and CEZO340 junctions calculated for ±4 V was found to be around 110 and 1700, respectively. This difference is mainly sourced from the current that flows through junctions in the reverse bias, as in CEZO340 structure it is over one order lower than in CEZO330. The I-V curves of both CEZO330 and CEZO340 samples exhibit nearly zero turn-on voltage (~0.2 V). A turn-on voltage $V_{on}$ = 0.4 V for ZnO/Si junction was previously obtained by Ghosh and Basak.[77] Similarly, CdO/Si structures are also known to exhibit such low $V_{on}$ values.[78–80] However, these examples generally suffer from poor rectification, particularly when compared to CEZO340 structure, which exhibits RF exceeding 50 at just 0.21V. An exceptionally low $V_{on}$, combined with a high RF, may be beneficial in terms of the use of such structures in current rectifiers operating at low voltages.

Nearly linear I-V characteristics for voltages over 0.2 V indicate a high series resistance of the junction.[81] This deviation from the ideal exponential I-V behavior is well-described by the single exponential model of a solar cell based on a p-n junction that takes into consideration both series resistance ($R_s$), and shunt resistance ($R_{sh}$):[82,83]

$$I = I_0 \left( e^{\frac{(V+IR_s)}{nV_{th}}} - 1 \right) + \frac{V+IR_s}{R_{sh}} - I_{ph}, \tag{6}$$

where $I_0$ is the saturation current, $V_{th}$ is the thermal voltage that equals to $k_B T/q$, where $q$ is the electric charge, $k_B$ is the Boltzmann constant, $T$ denotes the temperature and $n$ denotes the ideality factor, $I_{ph}$ is the photogenerated current. The exact solution of eq. (6) is expressed by[82]

$$I = \frac{V}{R_s+R_{sh}} + \frac{nV_{th}}{R_s} \cdot W\left\{ \frac{R_s I_0 R_{sh}}{nV_{th}(R_s+R_{sh})} \exp\left[ \frac{R_{sh}(R_s I_{ph} + R_s I_0 + V)}{nV_{th}(R_s+R_{sh})} \right] \right\} - \frac{R_{sh}(I_0 + I_{ph})}{R_s+R_{sh}}. \tag{7}$$

The junctions that did not exhibit issues related to Au/Zn(Cd)O Schottky contact formation were fitted using eq. (7) and OriginPro software to extract the electrical parameters. The results are presented in Figure 5e (for $I_{ph}$=0). In both cases, the shunt resistance was extremely high (> 1 MΩ), indicating the absence of relevant alternative current channels in the dark. The $R_s$ totaled 25.4 Ω for samples CEZO330 and 43.0 Ω for sample CEZO340. The impact of $R_s$ on device behavior is presented for the CEZO340 structure in the inset of Figure 5e. These curves were generated using eq. (7) with parameters extracted from the experimental data, followed by manual adjustment of $R_s$ values. The ideality factor, $n$, informs about current transport mechanisms within the junction. For sample CEZO330, the extracted value of $n$=1.5 lies within the range of 1-2, suggesting that current transport is governed by diffusion and generation-recombination processes.[84] In contrast, for sample CEZO340, $n$ exceeds 2, indicating that other mechanisms - such as carrier trapping or tunneling[84,85] - need to be considered.

All the samples generate photocurrent, as shown in Figure 6a and 6b. The responsivity of a detector is defined as:

$$R = \frac{I_{Light} - I_{Dark}}{P}, \tag{8}$$

where $I_{Light}$ is the current measured under illumination, $I_{Dark}$ denotes the current measured in the dark. These values were obtained without applying an external bias. $P$ presents the power of the incident light. As shown in Figure 6c, the extracted $R$ was measured locally rather than globally. The light spot had a square shape (0.85 mm × 0.85 mm) and was positioned to illuminate the Zn(Cd)O:Eu layer between the Au contacts, near the active Au contact connected to an Au wire. The structures convert photons with wavelengths longer than 370 nm, which corresponds to the 3.35 eV bandgap of ZnO. Photons of higher energy likely do not contribute to photocurrent due to strong absorption near the surface. The long-wavelength limit, around 1150 nm, aligns with the bandgap of Si. This makes the ZnCdO:Eu/Si system suitable for broadband light detection applications. Samples CEZO330 and CEZO340 exhibit the highest responsivity. For CEZO330, $R$ remains around 100 mW/A in the 450-1050 nm wavelength range. The responsivity of CEZO340 is even higher and closely follows the shape of the responsivity spectrum of the Si reference detector. The differences in $R$ values between these two samples cannot be explained by differences in optical reflection of the film, as the reflection spectra of both ZnCdO:Eu layers were nearly identical, as shown in Figure 6d. The oscillations observed in the measured spectra originate from thin-film interference.

Light and dark I-V curves were measured for the structures demonstrating the highest photocurrent generation (CEZO330 and CEZO340), as presented in Figure 6e and 6f. These measurements are combined with the calculated fill factor (FF) and open-circuit voltage ($V_{oc}$). The relatively low FF and $V_{oc}$ values, together with the steep character of light I-V curves in reverse bias, can mainly be explained by high series resistance.

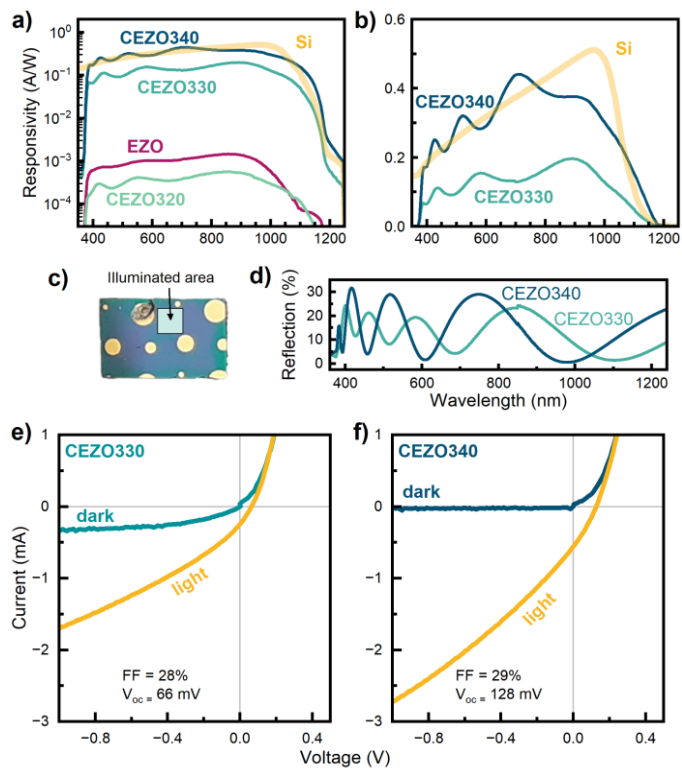

Figure 6. Photocurrent generation spectra of the investigated ZnCdO:Eu/Si junctions compared to a standard Si photodetector in a) semi-logarithmic and b) linear scale. Dark and light I-V characteristics of samples c) CEZO330 and d) CEZO340 measured with PET Solar Simulator (AM1.5G conditions, 1000W/m² light intensity, 25°C ambient temperature).

To estimate the response times of the samples, current-time (I-t) curves were measured, with the laser modulated at 50 ms intervals. Figure 7a presents the photocurrent response of sample CEZO330 to a 650 nm laser pulse at various temperatures. In addition to the current arising from the photovoltaic effect ($I_{ph}$), a signal attributed to the pyro-phototronic effect is also observed. When the laser is switched on, it induces localized heating of the structure. As the sample temperature, $T$, changes in time, $\frac{\partial T}{\partial t} \neq 0$, an additional electric field is induced within the ZnCdO:Eu layer. This results in a spike observed in the I-t characteristics, marked as '$I_{ph+pyro}$ heating'. Similarly, turning the laser off causes a decrease of the sample temperature and while the condition $\frac{\partial T}{\partial t} \neq 0$ is fulfilled, an electric field of opposite direction is induced, causing the flow of current marked as '$I_{ph+pyro}$ cooling'. The current oscillations observed under illumination are attributed to interference from the power grid. Both $I_{pyro-ph}$ and $I_{ph}$ change with temperature.

The behavior of $I_{pyro-ph}$, presented in Figure 7b, confirms that the observed spikes in the transients originate from the pyro-phototronic effect. As can be seen from the data, the $I_{pyro-ph}$ heating current decreases, while the $I_{pyro-ph}$ cooling current increases with temperature of the device. A higher sample temperature implies a higher difference of $T$ between the sample and the ambient environment ($RT$). Hence, after turning the laser off, the structure returns to its thermal equilibrium quicker at higher temperatures. For the same reason, the additional heat provided by the laser causes a smaller temperature rise at higher sample temperatures. This explains the observed decrease of $I_{pyro+ph}$ current. The reduction of $I_{ph}$ with increasing $T$ can be explained by the rise in the saturation current $I_0$, which is given by equation:[86]

$$I_0 = C \cdot T^3 \cdot \exp\left(-\frac{E_g}{k_B T}\right), \tag{9}$$

where $C$ is a constant, $E_g$ is a bandgap.

I-t transients measured under periodic 650 nm laser illumination for the studied samples are combined in Figure 7c. As expected from the photocurrent spectra, samples EZO and CEZO320 display almost negligible photocurrent generation, with only peaks related to the pyro-phototronic effect visible. In contrast, both the photovoltaic and pyro-phototronic effects occur in samples CEZO330 and CEZO340. Despite the total photocurrent values are similar (cf. Figure 6b), $I_{ph}$ generated in CEZO340 is significantly higher than in CEZO330. This is due to differences in the active device area.

Figure 7d shows zoomed-in I-t curves for samples CEZO330 and CEZO340 under 405 nm and 650 nm laser illumination, with rise ($\tau_r$) and fall ($\tau_f$) times estimated. The photocurrent was normalized to facilitate comparison between the samples. The rise time $\tau_r$ is defined as the time required for the photocurrent to increase from 10% to 90% of its maximum value, while, the fall time $\tau_f$ represents the time needed for the current to drop from 90% to 10%. For the 405 nm blue laser, $\tau_r$ and $\tau_f$ totaled 8 μs and 3 μs for sample CEZO330, 9 μs and 2 μs for CEZO340. Under 650 nm red laser illumination, $\tau_r$ and $\tau_f$ changed to 5 μs and 3 μs for CEZO330, and 3 μs and 2 μs for CEZO340. These measurements were performed using the maximum available laser powers of 251 mW/cm² for the 405 nm laser and 540 mW/cm² for the 650 nm laser. To determine whether light intensity has relevant impact on $\tau_r$ and $\tau_f$, additional tests were conducted on sample CEZO330, with light intensity modulated via optical filters. When illuminated with the 405 nm laser, $\tau_r$ and $\tau_f$ remained constant at 8 μs and 3 μs, respectively, across the 20-251 mW/cm² range. At a lower power of 9 mW/cm², $\tau_r$ increased to 14 μs and $\tau_f$ to 12 μs. Under even weaker illumination (0.7 mW/cm²), the pyro-phototronic effect was no longer evident - $\tau_r$ and $\tau_f$ increased to 120 μs and 110 μs, respectively, and no I$_{pyro+ph}$ spike was observed (cf. Figure 7e). In the case of the 650 nm laser, $\tau_r$ and $\tau_f$ stayed stable at 5 μs and 3 μs, respectively, across the 5-540 mW/cm² range. Only at the lowest measured light intensity of 0.75 mW/cm² did these times increase to 9 μs and 10 μs, respectively.

The ultrashort response times qualify the investigated structures as ultrafast photodetectors. Table 1 compares the response times of various self-powered PDs based on ZnO/Si heterojunctions reported in the literature. The devices presented in this work belong to the fastest ZnO/Si-based photodetectors with relatively high photocurrent generation, highlighting the potential of MBE-grown ZnCdO structures for the development of next-generation ultrafast PDs.

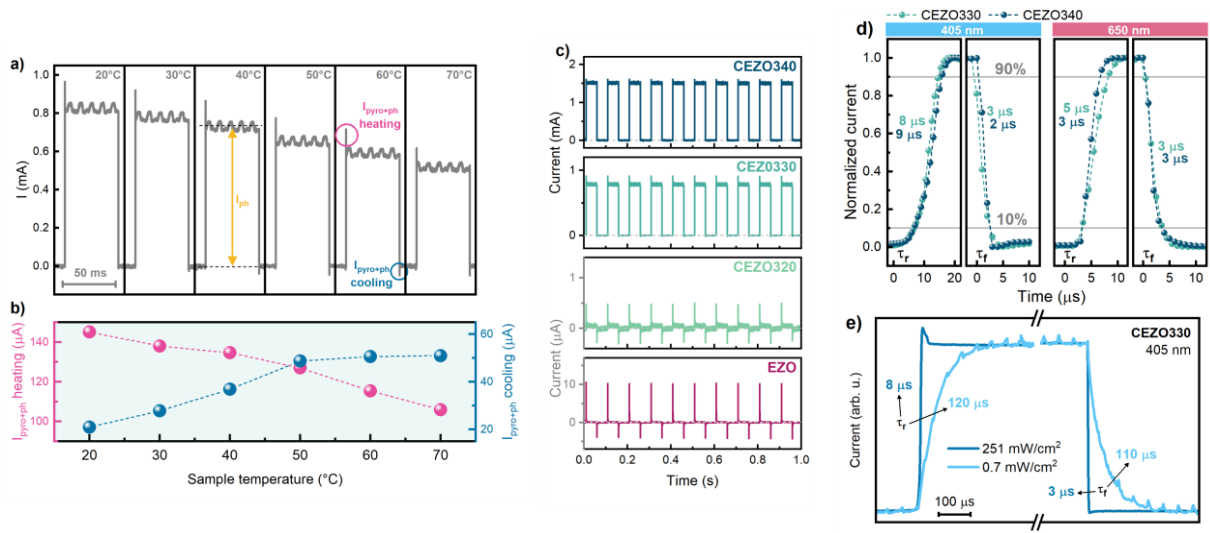

Figure 7. a) I-t chart of a single laser pulse measured for the CEZO330 structure at various temperatures (650 nm laser), b) mean values of the measured I$_{pyro+ph}$ during the laser switching on (heating) and off (cooling). c) Current-time curves of the investigated structures. Samples were illuminated with 650 nm laser for 50 ms every 100 ms. d) Zoomed-in I-t curves

illustrating the response times obtained for different wavelengths. e) Disappearance of the pyro-phototronic effect under low-intensity illumination conditions.

Table 1. Comparison of response times and photocurrent generation of self-biased PDs based solely on ZnO/Si heterojunctions, incorporating either undoped or doped ZnO. MW – microwire, NWs – nanowires, NRs – nanorods.

| Structure | Wavelength | Response time $\tau_r$ | Response time $\tau_f$ | Photocurrent (mA/W) | Reference |
|---|---|---|---|---|---|
| Ag/ZnO/Si/Ag | 365 nm | 1 s | 0.9 s | 5 | [75] |
| Ag/ZnO:Y NWs/Si/Al | 374 nm | 420 ms | 160 ms | Up to 200 | [87] |
| ITO/ZnMnO/Si/ITO | 900 nm | 3.4 ms | 4.1 ms | 100 | [88] |
| ITO/ZnO/Si/In | 405 nm | 130 µs | 120 µs | Up to 600 | [27] |
| ITO/ZnO NWs/Si/ITO | 940 nm | 113 µs | 200 µs | Up to 200 | [89] |
| ITO/ZnO:Ga MW/Si/Ni/Au | 370 nm | 79 µs | 132 µs | 200 | [90] |
| Al/ZnO NRs/Si/Al | 500 nm | 64 µs | 64 µs | 5 | [42] |
| ITO/ZnO NWs/Si/Cu | 1064 nm | 15 µs | 21 µs | 160 | [91] |
| ITO/ZnO/Si/Al | 850 nm | 10 µs | 31 µs | 6 | [41] |
| Au/ZnCdO:Eu/Si/Al | 405 nm | 8 µs | 3 µs | 70 | This work (CEZO330) |
| Au/ZnCdO:Eu/Si/Al | 650 nm | 5 µs | 3 µs | 135 | This work (CEZO330) |
| Au/ZnCdO:Eu/Si/Al | 405 nm | 9 µs | 2 µs | 185 | This work (CEZO340) |
| Au/ZnCdO:Eu/Si/Al | 650 nm | 3 µs | 3 µs | 370 | This work (CEZO340) |

**Conclusions**

This work reports Eu-doped ZnCdO random alloys grown on Si by molecular beam epitaxy. Presence of $Eu^{3+}$ ions in the structure was evidenced by µ-PL spectra. Eu and Cd, as foreign elements for the ZnO lattice can cause its distortion. As a result, defected crystal lattice was observed by means of cathodoluminescence and Raman spectra.

I-V curves shows that introduction of Cd into ZnO enables utilization of Au as an ohmic contact. Consequently, structures with the highest Cd concentration exhibit rectifying ratio at the level of 110 and 1700 measured at ±4 V. Moreover, these samples display high responsivity in a broad spectral range 370-1150 nm, reaching over 400 mA/W for 700 nm without any external electrical bias, making the presented PDs attractive for energy-saving optoelectronics. Pyro-phototronic effect led to ultrafast reaction times of the PDs. The fastest detection was recorded for the sample with the highest Cd concentration illuminated by 650 nm light – 3 µs for both rise and fall time. These result belong to the shortest detection times presented in the literature for self-powered PDs based solely on ZnO/Si junction. Furthermore, the ultrafast detection were stable over a wide range of light intensities. All the described features allow utilization of the proposed structures to detect fast-changing signals under various light conditions (when it comes to both wavelength and intensity) without the need of using any power source.

This study presents a novel approach to revealing the interaction between Cd and Eu doping in ZnO layers and the pyro-phototronic effect, offering valuable insights for improving the response of ZnO-based self-powered photodetectors in fast photosensing applications.

## Author Contributions

I.P. – conceptualization, formal analysis, investigation, writing – original draft, visualization; E.Z. – investigation, writing – review & editing, supervision; A.W. – investigation, writing – review & editing; A.L. – resources, investigation; R.J. – investigation; E.P. – writing – review & editing, validation, supervision.

## Declaration of Competing Interest

The authors declare that they have no known competing financial interests or personal relationships that could have appeared to influence the work reported in this paper.

## Data availability

Data will be made available on request.

## Supporting Information

SIMS data measured for CEZO340 sample (DOCX).

## Acknowledgements

This work was supported by the Polish National Science Center, grant no. 2021/41/B/ST5/00216.

## References

(1) Zúñiga-Pérez, J. ZnCdO: Status after 20 Years of Research. *Mater Sci Semicond Process* **2017**, *69*, 36–43. https://doi.org/10.1016/j.mssp.2016.12.002.

(2) Özgür, Ü.; Alivov, Ya. I.; Liu, C.; Teke, A.; Reshchikov, M. A.; Doğan, S.; Avrutin, V.; Cho, S.-J.; Morkoç, H. A Comprehensive Review of ZnO Materials and Devices. *J Appl Phys* **2005**, *98* (4). https://doi.org/10.1063/1.1992666.

(3)  Vasheghani Farahani, S. K.; Muñoz-Sanjosé, V.; Zúñiga-Pérez, J.; McConville, C. F.; Veal, T. D. Temperature Dependence of the Direct Bandgap and Transport Properties of CdO. *Appl Phys Lett* **2013**, *102* (2). https://doi.org/10.1063/1.4775691.

(4)  Pietrzyk, M. A.; Wierzbicka, A.; Zielony, E.; Pieniazek, A.; Szymon, R.; Placzek-Popko, E. Fundamental Studies of ZnO Nanowires with ZnCdO/ZnO Multiple Quantum Wells Grown for Tunable Light Emitters. *Sens Actuators A Phys* **2020**, *315*, 112305. https://doi.org/10.1016/j.sna.2020.112305.

(5)  Venkatachalapathy, V.; Galeckas, A.; Sellappan, R.; Chakarov, D.; Kuznetsov, A. Yu. Tuning Light Absorption by Band Gap Engineering in ZnCdO as a Function of MOVPE-Synthesis Conditions and Annealing. *J Cryst Growth* **2011**, *315* (1), 301–304. https://doi.org/10.1016/j.jcrysgro.2010.09.056.

(6)  Jang, H.; Saito, K.; Guo, Q.; Tanaka, T.; Yu, K. M.; Walukiewicz, W. Growth of Low Resistive Al-Doped ZnCdO Thin Films with Rocksalt Structure for Transparent Conductive Oxide Thin Films. In *2020 47th IEEE Photovoltaic Specialists Conference (PVSC)*; IEEE, 2020; pp 0191–0193. https://doi.org/10.1109/PVSC45281.2020.9300563.

(7)  Kalusniak, S.; Sadofev, S.; Puls, J.; Henneberger, F. ZnCdO/ZnO – a New Heterosystem for Green-wavelength Semiconductor Lasing. *Laser Photon Rev* **2009**, *3* (3), 233–242. https://doi.org/10.1002/lpor.200810040.

(8)  Pearton, S. J.; Ren, F. P-Type Doping of ZnO Films and Growth of Tenary ZnMgO and ZnCdO: Application to Light Emitting Diodes and Laser Diodes. *International Materials Reviews* **2014**, *59* (2), 61–83. https://doi.org/10.1179/1743280413Y.0000000025.

(9)  Yamamoto, K.; Nakamura, A.; Temmyo, J.; Munoz, E.; Hierro, A. Green Electroluminescence From ZnCdO Multiple Quantum-Well Light-Emitting Diodes Grown by Remote-Plasma-Enhanced Metal–Organic Chemical Vapor Deposition. *IEEE Photonics Technology Letters* **2011**, *23* (15), 1052–1054. https://doi.org/10.1109/LPT.2011.2152389.

(10) Kim, J.-R. Hole Transport Improvement in CdZnO/ZnO Light Emitting Diodes with Wedge Shaped Electron Blocking Layers. *Journal of the Korean Physical Society* **2023**, *82* (10), 981–984. https://doi.org/10.1007/s40042-023-00814-w.

(11) Stachowicz, M.; Sajkowski, J. M.; Pietrzyk, M. A.; Faye, D. Nd.; Magalhaes, S.; Alves, E.; Reszka, A.; Pieniążek, A.; Kozanecki, A. Investigation of Interdiffusion in Thin Films of ZnO/ZnCdO Grown by Molecular Beam Epitaxy. *Thin Solid Films* **2023**, *781*, 140003. https://doi.org/10.1016/j.tsf.2023.140003.

(12) Lysak, A.; Przeździecka, E.; Wierzbicka, A.; Dłużewski, P.; Sajkowski, J.; Morawiec, K.; Kozanecki, A. The Influence of the Growth Temperature on the Structural

Properties of {CdO/ZnO}$_{30}$ Superlattices. *Cryst Growth Des* **2023**, *23* (1), 134–141. https://doi.org/10.1021/acs.cgd.2c00826.

(13) Przeździecka, E.; Wierzbicka, A.; Lysak, A.; Dłużewski, P.; Adhikari, A.; Sybilski, P.; Morawiec, K.; Kozanecki, A. Nanoscale Morphology of Short-Period {CdO/ZnO} Superlattices Grown by MBE. *Cryst Growth Des* **2022**, *22* (2), 1110–1115. https://doi.org/10.1021/acs.cgd.1c01065.

(14) Przeździecka, E.; Lysak, A.; Adhikari, A.; Stachowicz, M.; Wierzbicka, A.; Jakiela, R.; Khosravizadeh, Z.; Sybilski, P.; Kozanecki, A. Influence of the Growth Temperature and Annealing on the Optical Properties of {CdO/ZnO}30 Superlattices. *J Lumin* **2024**, *269*, 120481. https://doi.org/10.1016/j.jlumin.2024.120481.

(15) Lysak, A.; Przeździecka, E.; Jakiela, R.; Reszka, A.; Witkowski, B.; Khosravizadeh, Z.; Adhikari, A.; Sajkowski, J. M.; Kozanecki, A. Effect of Rapid Thermal Annealing on Short Period {CdO/ZnO}m SLs Grown on m-Al2O3. *Mater Sci Semicond Process* **2022**, *142*, 106493. https://doi.org/10.1016/j.mssp.2022.106493.

(16) Seshu Kumar, M.; Jayaprada, P.; Ravikumar, R. V. S. S. N.; Rao, M. C. Structural and Optical Studies on Ti-Doped ZnCdO Nanocomposites for Optoelectronic Device Application. *J Appl Spectrosc* **2023**, *90* (4), 860–866. https://doi.org/10.1007/s10812-023-01607-6.

(17) Parthasaradi, V.; Kavitha, M.; Sridevi, A.; Rubia, J. J. Novel Rare-Earth Eu and La Co-Doped ZnO Nanoparticles Synthesized via Co-Precipitation Method: Optical, Electrical, and Magnetic Properties. *Journal of Materials Science: Materials in Electronics* **2022**, *33* (34), 25805–25819. https://doi.org/10.1007/s10854-022-09272-9.

(18) Kumawat, A.; Misra, K. P.; Chattopadhyay, S. Band Gap Engineering and Relationship with Luminescence in Rare-Earth Elements Doped ZnO: An Overview. *Materials Technology* **2022**, *37* (11), 1595–1610. https://doi.org/10.1080/10667857.2022.2082351.

(19) Perlikowski, I.; Zielony, E.; Lysak, A.; Jakieła, R.; Przeździecka, E. Manifestation of Eu Dopants in Raman Spectra and Doping Concentration Profiles of {ZnCdO/ZnO} Superlattices. *Cryst Growth Des* **2024**, *24* (16), 6691–6700. https://doi.org/10.1021/acs.cgd.4c00619.

(20) Üzar, N. Investigation of Detailed Physical Properties and Solar Cell Performances of Various Type Rare Earth Elements Doped ZnO Thin Films. *Journal of Materials Science: Materials in Electronics* **2018**, *29* (12), 10471–10479. https://doi.org/10.1007/s10854-018-9111-3.

(21) Honglin, L.; Yingbo, L.; Jinzhu, L.; Ke, Y. Experimental and First-Principles Studies of Structural and Optical Properties of Rare Earth (RE = La, Er, Nd) Doped ZnO. *J Alloys Compd* **2014**, *617*, 102–107. https://doi.org/10.1016/j.jallcom.2014.08.019.

(22) Krajewski, T. A.; Ratajczak, R.; Kobyakov, S.; Wozniak, W.; Kopalko, K.; Guziewicz, E. Electrical Properties of ZnO Films Implanted with Rare Earth and Their Relationship with Structural and Optical Parameters. *Materials Science and Engineering: B* **2022**, *275*, 115526. https://doi.org/10.1016/j.mseb.2021.115526.

(23) Stojadinović, S.; Ćirić, A. Photoluminescence of ZnO:Eu3+ and ZnO:Tb3+ Coatings Formed by Plasma Electrolytic Oxidation of Pure Zinc Substrate. *J Lumin* **2021**, *235*, 118022. https://doi.org/10.1016/j.jlumin.2021.118022.

(24) Tan, Y.; Fang, Z.; Chen, W.; He, P. Structural, Optical and Magnetic Properties of Eu-Doped ZnO Films. *J Alloys Compd* **2011**, *509* (21), 6321–6324. https://doi.org/10.1016/j.jallcom.2011.03.084.

(25) Hasabeldaim, E.; Ntwaeaborwa, O. M.; Kroon, R. E.; Swart, H. C. Structural, Optical and Photoluminescence Properties of Eu Doped ZnO Thin Films Prepared by Spin Coating. *J Mol Struct* **2019**, *1192*, 105–114. https://doi.org/10.1016/j.molstruc.2019.04.128.

(26) Swapna, R.; SrinivasaReddy, T.; Venkateswarlu, K.; Kumar, M. C. S. Effect of Post-Annealing on the Properties of Eu Doped ZnO Nano Thin Films. *Procedia Materials Science* **2015**, *10*, 723–729. https://doi.org/10.1016/j.mspro.2015.06.085.

(27) Yu, W.; Hao, L.; Guo, F.; Zhang, M.; Li, S.; Hu, B.; Zhang, B.; Liu, Y. High-Performance Broadband Si/ZnO Heterojunction Photodetector Based on Pyro-Phototronic Effect. *Opt Mater (Amst)* **2024**, *147*, 114752. https://doi.org/10.1016/j.optmat.2023.114752.

(28) Xu, Z.; Zhang, Y.; Wang, Z. ZnO-Based Photodetector: From Photon Detector to Pyro-Phototronic Effect Enhanced Detector. *J Phys D Appl Phys* **2019**, *52* (22), 223001. https://doi.org/10.1088/1361-6463/ab0728.

(29) Li, Q.; Huang, J.; Meng, J.; Li, Z. Enhanced Performance of a Self-Powered ZnO Photodetector by Coupling LSPR-Inspired Pyro-Phototronic Effect and Piezo-Phototronic Effect. *Adv Opt Mater* **2022**, *10* (7). https://doi.org/10.1002/adom.202102468.

(30) Yin, B.; Zhang, H.; Qiu, Y.; Luo, Y.; Zhao, Y.; Hu, L. The Light-Induced Pyro-Phototronic Effect Improving a ZnO/NiO/Si Heterojunction Photodetector for Selectively Detecting Ultraviolet or Visible Illumination. *Nanoscale* **2017**, *9* (44), 17199–17206. https://doi.org/10.1039/C7NR06037H.

(31) Wang, Z.; Yu, R.; Wang, X.; Wu, W.; Wang, Z. L. Ultrafast Response P-Si/N-ZnO Heterojunction Ultraviolet Detector Based on Pyro-Phototronic Effect. *Advanced Materials* **2016**, *28* (32), 6880–6886. https://doi.org/10.1002/adma.201600884.

(32) Goel, V.; Kumar, Y.; Rawat, G.; Kumar, H. Self-Powered Photodetectors: A Device Engineering Perspective. *Nanoscale* **2024**, *16* (19), 9235–9258. https://doi.org/10.1039/D4NR00607K.

(33) Peng, W.; Yu, R.; Wang, X.; Wang, Z.; Zou, H.; He, Y.; Wang, Z. L. Temperature Dependence of Pyro-Phototronic Effect on Self-Powered ZnO/Perovskite Heterostructured Photodetectors. *Nano Res* **2016**, *9* (12), 3695–3704. https://doi.org/10.1007/s12274-016-1240-5.

(34) Wan, P.; Jiang, M.; Xu, T.; Liu, Y.; Fang, X.; Kan, C. Doping Concentration Influenced Pyro-Phototronic Effect in Self-Powered Photodetector Based on Ga-Incorporated ZnO Microwire/p$^+$-GaN Heterojunction. *Adv Opt Mater* **2022**, *10* (2). https://doi.org/10.1002/adom.202101851.

(35) Veeralingam, S.; Yadav, P.; Badhulika, S. An Fe-Doped ZnO/BiVO$_4$ Heterostructure-Based Large Area, Flexible, High-Performance Broadband Photodetector with an Ultrahigh Quantum Yield. *Nanoscale* **2020**, *12* (16), 9152–9161. https://doi.org/10.1039/C9NR10776B.

(36) Deka Boruah, B.; Naidu Majji, S.; Nandi, S.; Misra, A. Doping Controlled Pyro-Phototronic Effect in Self-Powered Zinc Oxide Photodetector for Enhancement of Photoresponse. *Nanoscale* **2018**, *10* (7), 3451–3459. https://doi.org/10.1039/C7NR08125A.

(37) Yatskiv, R.; Grym, J.; Bašinová, N.; Kučerová, Š.; Vaniš, J.; Piliai, L.; Vorokhta, M.; Veselý, J.; Maixner, J. Defect-Mediated Energy Transfer in ZnO Thin Films Doped with Rare-Earth Ions. *J Lumin* **2023**, *253*, 119462. https://doi.org/10.1016/j.jlumin.2022.119462.

(38) Ratajczak, R.; Guziewicz, E.; Prucnal, S.; Mieszczynski, C.; Jozwik, P.; Barlak, M.; Romaniuk, S.; Gieraltowska, S.; Wozniak, W.; Heller, R.; Kentsch, U.; Facsko, S. Enhanced Luminescence of Yb3+ Ions Implanted to ZnO through the Selection of Optimal Implantation and Annealing Conditions. *Materials* **2023**, *16* (5), 1756. https://doi.org/10.3390/ma16051756.

(39) Krajewski, T. A.; Ratajczak, R.; Kobyakov, S.; Wozniak, W.; Kopalko, K.; Guziewicz, E. Electrical Properties of ZnO Films Implanted with Rare Earth and Their Relationship with Structural and Optical Parameters. *Materials Science and Engineering: B* **2022**, *275*, 115526. https://doi.org/10.1016/j.mseb.2021.115526.

(40) Röder, R.; Geburt, S.; Zapf, M.; Franke, D.; Lorke, M.; Frauenheim, T.; da Rosa, A. L.; Ronning, C. Transition Metal and Rare Earth Element Doped Zinc Oxide Nanowires for Optoelectronics. *physica status solidi (b)* **2019**, *256* (4). https://doi.org/10.1002/pssb.201800604.

(41) Ahmed, A. A.; Qahtan, T. F.; Hashim, M. R.; Nomaan, A. T.; Al-Hardan, N. H.; Rashid, M. Eco-Friendly Ultrafast Self-Powered p-Si/n-ZnO Photodetector Enhanced by Photovoltaic-Pyroelectric Coupling Effect. *Ceram Int* **2022**, *48* (11), 16142–16155. https://doi.org/10.1016/j.ceramint.2022.02.162.

(42) Mondal, S.; Halder, S.; Basak, D. Ultrafast and Ultrabroadband UV–Vis-NIR Photosensitivity under Reverse and Self-Bias Conditions by n$^+$-ZnO/n-Si Isotype Heterojunction with >1 KHz Bandwidth. *ACS Appl Electron Mater* **2023**, *5* (2), 1212–1223. https://doi.org/10.1021/acsaelm.2c01668.

(43) Kaim, A.; Gwóźdź, K. MyDLTS: LabVIEW Based Software for Deep Level Transient Spectroscopy Using Impedance Analyser. *SoftwareX* **2024**, *26*, 101679. https://doi.org/10.1016/j.softx.2024.101679.

(44) Lysak, A.; Przeździecka, E.; Wierzbicka, A.; Jakiela, R.; Khosravizadeh, Z.; Szot, M.; Adhikari, A.; Kozanecki, A. Temperature Dependence of the Bandgap of Eu Doped {ZnCdO/ZnO}30 Multilayer Structures. *Thin Solid Films* **2023**, *781*, 139982. https://doi.org/10.1016/j.tsf.2023.139982.

(45) Heng, C. L.; Xiang, W.; Su, W. Y.; Gao, Y. K.; Yin, P. G.; Finstad, T. G. Effect of Eu Doping on the near Band Edge Emission of Eu Doped ZnO Thin Films after High Temperature Annealing. *J Lumin* **2019**, *210*, 363–370. https://doi.org/10.1016/j.jlumin.2019.02.062.

(46) Lupan, O.; Pauporté, T.; Viana, B.; Aschehoug, P.; Ahmadi, M.; Cuenya, B. R.; Rudzevich, Y.; Lin, Y.; Chow, L. Eu-Doped ZnO Nanowire Arrays Grown by Electrodeposition. *Appl Surf Sci* **2013**, *282*, 782–788. https://doi.org/10.1016/j.apsusc.2013.06.053.

(47) Hasabeldaim, E.; Ntwaeaborwa, O. M.; Kroon, R. E.; Swart, H. C. Structural, Optical and Photoluminescence Properties of Eu Doped ZnO Thin Films Prepared by Spin Coating. *J Mol Struct* **2019**, *1192*, 105–114. https://doi.org/10.1016/j.molstruc.2019.04.128.

(48) Singh, A.; Arya, P.; Choudhary, D.; Kumar, S.; Srivastava, A. K.; Singh, I. B. Cost-Effective ZnO–Eu3+ Films with Efficient Energy Transfer between Host and Dopant. *SN Appl Sci* **2020**, *2* (5), 870. https://doi.org/10.1007/s42452-020-2670-y.

(49) Tan, Y.; Fang, Z.; Chen, W.; He, P. Structural, Optical and Magnetic Properties of Eu-Doped ZnO Films. *J Alloys Compd* **2011**, *509* (21), 6321–6324. https://doi.org/10.1016/j.jallcom.2011.03.084.

(50) Wierzbicka, A.; Tchutchulashvili, G.; Sobanska, M.; Klosek, K.; Minikayev, R.; Domagala, J. Z.; Borysiuk, J.; Zytkiewicz, Z. R. Arrangement of GaN Nanowires on Si(001) Substrates Studied by X-Ray Diffraction: Importance of Silicon Nitride Interlayer. *Appl Surf Sci* **2017**, *425*, 1014–1019. https://doi.org/10.1016/j.apsusc.2017.07.075.

(51) Renninger, M. "Umweganregung", Eine Bisher Unbeachtete Wechselwirkungserscheinung Bei Raumgitterinterferenzen. *Zeitschrift fur Physik* **1937**, *106* (3–4), 141–176. https://doi.org/10.1007/BF01340315.

(52) Bindu, P.; Thomas, S. Estimation of Lattice Strain in ZnO Nanoparticles: X-Ray Peak Profile Analysis. *Journal of Theoretical and Applied Physics* **2014**, *8* (4), 123–134. https://doi.org/10.1007/s40094-014-0141-9.

(53) Adhikari, A.; Wierzbicka, A.; Lysak, A.; Sybilski, P.; Witkowski, B. S.; Przeździecka, E. Effective Mg Incorporation in CdMgO Alloy on Quartz Substrate Grown by Plasma-Assisted MBE. *Acta Phys Pol A* **2023**, *143* (3), 228–237. https://doi.org/10.12693/APhysPolA.143.228.

(54) Mustapha, S.; Ndamitso, M. M.; Abdulkareem, A. S.; Tijani, J. O.; Shuaib, D. T.; Mohammed, A. K.; Sumaila, A. Comparative Study of Crystallite Size Using Williamson-Hall and Debye-Scherrer Plots for ZnO Nanoparticles. *Advances in Natural Sciences: Nanoscience and Nanotechnology* **2019**, *10* (4), 045013. https://doi.org/10.1088/2043-6254/ab52f7.

(55) Dakhel, A. A. Bandgap Narrowing in CdO Doped with Europium. *Opt Mater (Amst)* **2009**, *31* (4), 691–695. https://doi.org/10.1016/j.optmat.2008.08.001.

(56) Huang, X.; Zhou, Y.; Wu, W.; Xu, J.; Liu, S.; Liu, D.; Wu, J. Effect of $Zn^{2+}$ Substitution on the Structure and Magnetic Properties of $Co_{0.5}Cu_{0.5}Fe_2O_4$ Synthesized by Solvothermal Method. *J Electron Mater* **2016**, *45* (6), 3113–3120. https://doi.org/10.1007/s11664-016-4400-1.

(57) Liebhaber, M.; Bass, U.; Bayersdorfer, P.; Geurts, J.; Speiser, E.; Räthel, J.; Baumann, A.; Chandola, S.; Esser, N. Surface Phonons of the Si(111)-(7×7) Reconstruction Observed by Raman Spectroscopy. *Phys Rev B* **2014**, *89* (4), 045313. https://doi.org/10.1103/PhysRevB.89.045313.

(58) Zielony, E.; Szalewska, G.; Pietrzyk, M. A. Probing N-ZnMgO/p-Si Nanowire Junctions: Insights into Composition, Strain, and Defects via Raman Spectroscopy and Electrical Measurements. *J Alloys Compd* **2025**, *1010*, 177851. https://doi.org/10.1016/j.jallcom.2024.177851.

(59) Borkovska, L. V.; Khomenkova, L.; Korsunska, O.; Kolomys, O.; Strelchuk, V.; Sabov, T.; Venger, E.; Kryshtab, T.; Melnichuk, O.; Melnichuk, L.; Guillaume, C.; Labbe, C.; Portier, X. Influence of Annealing on Luminescence and Energy Transfer in ZnO Multilayer Structure Co-Doped with Tb and Eu. *Thin Solid Films* **2019**, *692*, 137634. https://doi.org/10.1016/j.tsf.2019.137634.

(60) Mondal, A.; Pal, S.; Sarkar, A.; Bhattacharya, T. S.; Pal, S.; Singha, A.; Ray, S. K.; Kumar, P.; Kanjilal, D.; Jana, D. Raman Investigation of N-implanted ZnO: Defects, Disorder and Recovery. *Journal of Raman Spectroscopy* **2019**, *50* (12), 1926–1937. https://doi.org/10.1002/jrs.5732.

(61) Song, Y.; Zhang, S.; Zhang, C.; Yang, Y.; Lv, K. Raman Spectra and Microstructure of Zinc Oxide Irradiated with Swift Heavy Ion. *Crystals (Basel)* **2019**, *9* (8), 395. https://doi.org/10.3390/cryst9080395.

(62) Castro, T. J.; Franco, A.; Pessoni, H. V. S.; Rodrigues, P. A. M.; Morais, P. C.; da Silva, S. W. Investigation of Additional Raman Modes in ZnO and Eu0.01Zn0.99O Nanoparticles Synthesized by the Solution Combustion Method. *J Alloys Compd* **2017**, *691*, 416–421. https://doi.org/10.1016/j.jallcom.2016.08.297.

(63) Manjón, F. J.; Marí, B.; Serrano, J.; Romero, A. H. Silent Raman Modes in Zinc Oxide and Related Nitrides. *J Appl Phys* **2005**, *97* (5). https://doi.org/10.1063/1.1856222.

(64) Kolomys, O.; Romanyuk, A.; Strelchuk, V.; Lashkarev, G.; Khyzhun, O.; Timofeeva, I.; Lazorenko, V.; Khomyak, V. Optical and Structural Studies of Phase Transformations and Composition Fluctuations at Annealing of Zn1-XCdxO Films Grown by Dc Magnetron Sputtering. *Semiconductor Physics Quantum Electronics and Optoelectronics* **2014**, *17* (3), 275–283. https://doi.org/10.15407/spqeo17.03.275.

(65) Meyer, B. K.; Sann, J.; Lautenschläger, S.; Wagner, M. R.; Hoffmann, A. Ionized and Neutral Donor-Bound Excitons in ZnO. *Phys Rev B* **2007**, *76* (18), 184120. https://doi.org/10.1103/PhysRevB.76.184120.

(66) Hasabeldaim, E. H. H.; Ntwaeaborwa, O. M.; Kroon, R. E.; Coetsee, E.; Swart, H. C. Luminescence Properties of Eu Doped ZnO PLD Thin Films: The Effect of Oxygen Partial Pressure. *Superlattices Microstruct* **2020**, *139*, 106432. https://doi.org/10.1016/j.spmi.2020.106432.

(67) Zhang, Y.; Liu, Y.; Wu, L.; Xie, E.; Chen, J. Photoluminescence and ZnO → Eu$^{3+}$ Energy Transfer in Eu$^{3+}$-Doped ZnO Nanospheres. *J Phys D Appl Phys* **2009**, *42* (8), 085106. https://doi.org/10.1088/0022-3727/42/8/085106.

(68) Lysak, A.; Wierzbicka, A.; Magalhaes, S.; Dłużewski, P.; Jakieła, R.; Szot, M.; Khosravizadeh, Z.; Adhikari, A.; Kozanecki, A.; Przeździecka, E. Structural and Optical Properties of in Situ Eu-Doped ZnCdO/ZnMgO Superlattices Grown by Plasma-Assisted Molecular Beam Epitaxy. *Nanoscale* **2025**, *17* (12), 7055–7065. https://doi.org/10.1039/D4NR04847D.

(69) Lu, M.-Y.; Lu, M.-P.; You, S.-J.; Chen, C.-W.; Wang, Y.-J. Quantifying the Barrier Lowering of ZnO Schottky Nanodevices under UV Light. *Sci Rep* **2015**, *5* (1), 15123. https://doi.org/10.1038/srep15123.

(70) Pandis, Ch.; Brilis, N.; Bourithis, E.; Tsamakis, D.; Ali, H.; Krishnamoorthy, S.; Iliadis, A. A.; Kompitsas, M. Low–Temperature Hydrogen Sensors Based on Au Nanoclusters and Schottky Contacts on ZnO Films Deposited by Pulsed Laser Deposition on Si and ${\hbox{SiO}}_{2}$ Substrates. *IEEE Sens J* **2007**, *7* (3), 448–454. https://doi.org/10.1109/JSEN.2007.891944.

(71) Szymon, R.; Zielony, E.; Lysak, A.; Pietrzyk, M. A. Influence of the Type of Interlayer on Current Transport Mechanisms and Defects in N-ZnO/ZnCdO/p-Si and n-ZnCdO/ZnO/p-Si Heterojunctions Grown by Molecular Beam Epitaxy. *J Alloys Compd* **2023**, *951*, 169859. https://doi.org/10.1016/j.jallcom.2023.169859.

(72) Lu, Y.; Huang, J.; Li, B.; Tang, K.; Ma, Y.; Cao, M.; Wang, L.; Wang, L. A Boron and Gallium Co-Doped ZnO Intermediate Layer for ZnO/Si Heterojunction Diodes. *Appl Surf Sci* **2018**, *428*, 61–65. https://doi.org/10.1016/j.apsusc.2017.09.053.

(73) Dutta, M.; Basak, D. P - Zn O ∕ n - Si Heterojunction: Sol-Gel Fabrication, Photoresponse Properties, and Transport Mechanism. *Appl Phys Lett* **2008**, *92* (21). https://doi.org/10.1063/1.2937124.

(74) Cao, T.; Luo, L.; Huang, Y.; Ye, B.; She, J.; Deng, S.; Chen, J.; Xu, N. Integrated ZnO Nano-Electron-Emitter with Self-Modulated Parasitic Tunneling Field Effect Transistor at the Surface of the p-Si/ZnO Junction. *Sci Rep* **2016**, *6* (1), 33983. https://doi.org/10.1038/srep33983.

(75) Al-Hardan, N. H.; Mohd Rashid, M. M.; Abdul Aziz, A.; Ahmed, N. M. Low Power Consumption UV Sensor Based on N-ZnO/p-Si Junctions. *Journal of Materials Science: Materials in Electronics* **2019**, *30* (21), 19639–19646. https://doi.org/10.1007/s10854-019-02337-2.

(76) Jensen, I. J. T.; Johansen, K. M.; Zhan, W.; Venkatachalapathy, V.; Brillson, L.; Kuznetsov, A. Yu.; Prytz, Ø. Bandgap and Band Edge Positions in Compositionally Graded ZnCdO. *J Appl Phys* **2018**, *124* (1). https://doi.org/10.1063/1.5036710.

(77) Ghosh, R.; Basak, D. Electrical and Ultraviolet Photoresponse Properties of Quasialigned ZnO Nanowires/p-Si Heterojunction. *Appl Phys Lett* **2007**, *90* (24). https://doi.org/10.1063/1.2748333.

(78) Ravikumar, M.; Ganesh, V.; Shkir, M.; Chandramohan, R.; Arun Kumar, K. D.; Valanarasu, S.; Kathalingam, A.; AlFaify, S. Fabrication of Eu Doped CdO [Al/Eu-NCdO/p-Si/Al] Photodiodes by Perfume Atomizer Based Spray Technique for Opto-Electronic Applications. *J Mol Struct* **2018**, *1160*, 311–318. https://doi.org/10.1016/j.molstruc.2018.01.095.

(79) Ismail, R. A.; Al-Samarai, A.-M. E.; Mohmed, S. J.; Ahmed, H. H. Characteristics of Nanostructured CdO/Si Heterojunction Photodetector Synthesized by CBD. *Solid State Electron* **2013**, *82*, 115–121. https://doi.org/10.1016/j.sse.2013.02.035.

(80) Hamadi, O. A. Characteristics of CdO—Si Heterostructure Produced by Plasma-Induced Bonding Technique. *Proceedings of the Institution of Mechanical Engineers, Part L: Journal of Materials: Design and Applications* **2008**, *222* (1), 65–72. https://doi.org/10.1243/14644207JMDA56.

(81) Jain, A. Exact Analytical Solutions of the Parameters of Real Solar Cells Using Lambert W-Function. *Solar Energy Materials and Solar Cells* **2004**, *81* (2), 269–277. https://doi.org/10.1016/j.solmat.2003.11.018.

(82) Jain, A.; Kapoor, A. Exact Analytical Solutions of the Parameters of Real Solar Cells Using Lambert W-Function. *Solar Energy Materials and Solar Cells* **2004**, *81* (2), 269–277. https://doi.org/10.1016/j.solmat.2003.11.018.

(83) Carrero, C.; Rodríguez, J.; Ramírez, D.; Platero, C. Simple Estimation of PV Modules Loss Resistances for Low Error Modelling. *Renew Energy* **2010**, *35* (5), 1103–1108. https://doi.org/10.1016/j.renene.2009.10.025.

(84) Płaczek-Popko, E.; Paradowska, K. M.; Pietrzyk, M. A.; Kozanecki, A. Carrier Transport Mechanisms in the ZnO Based Heterojunctions Grown by MBE. *Opto-Electronics Review* **2017**, *25* (3), 181–187. https://doi.org/10.1016/j.opelre.2017.06.010.

(85) Zielony, E.; Płaczek-Popko, E.; Nowakowski, P.; Gumienny, Z.; Suchocki, A.; Karczewski, G. Electro-Optical Characterization of Ti/Au–ZnTe Schottky Diodes with CdTe Quantum Dots. *Mater Chem Phys* **2012**, *134* (2–3), 821–828. https://doi.org/10.1016/j.matchemphys.2012.03.075.

(86) Singh, P.; Ravindra, N. M. Temperature Dependence of Solar Cell Performance—an Analysis. *Solar Energy Materials and Solar Cells* **2012**, *101*, 36–45. https://doi.org/10.1016/j.solmat.2012.02.019.

(87) Saha, R.; Dalapati, G. K.; Chakrabarti, S.; Karmakar, A.; Chattopadhyay, S. Yttrium (Y) Doped ZnO Nanowire/p-Si Heterojunction Devices for Efficient Self-Powered UV-Sensing Applications. *Vacuum* **2022**, *202*, 111214. https://doi.org/10.1016/j.vacuum.2022.111214.

(88) Wang, L.; Xue, H.; Zhu, M.; Gao, Y.; Wang, Z. Graded Strain-Enhanced Pyro-Phototronic Photodetector with a Broad and Plateau Band. *Nano Energy* **2022**, *97*, 107163. https://doi.org/10.1016/j.nanoen.2022.107163.

(89) Wang, Y.; Zhu, Y.; Gu, H.; Wang, X. Enhanced Performances of N-ZnO Nanowires/p-Si Heterojunctioned Pyroelectric Near–Infrared Photodetectors via the Plasmonic Effect. *ACS Appl Mater Interfaces* **2021**, *13* (48), 57750–57758. https://doi.org/10.1021/acsami.1c14319.

(90) Dai, R.; Liu, Y.; Wu, J.; Wan, P.; Zhu, X.; Kan, C.; Jiang, M. Self-Powered Ultraviolet Photodetector Based on an n-ZnO:Ga Microwire/p-Si Heterojunction with the Performance Enhanced by a Pyro-Phototronic Effect. *Opt Express* **2021**, *29* (19), 30244. https://doi.org/10.1364/OE.439587.

(91) Chen, L.; Wang, B.; Dong, J.; Gao, F.; Zheng, H.; He, M.; Wang, X. Insights into the Pyro-Phototronic Effect in p-Si/n-ZnO Nanowires Heterojunction toward High-

Performance near-Infrared Photosensing. *Nano Energy* **2020**, *78*, 105260. https://doi.org/10.1016/j.nanoen.2020.105260.